\documentclass[conference]{IEEEtran}
\IEEEoverridecommandlockouts
\usepackage{cite}
\usepackage{amsmath,amssymb,amsfonts}
\usepackage{algorithmic}
\usepackage{graphicx}
\usepackage{textcomp}
\usepackage{xcolor}
\usepackage{comment}
\def\BibTeX{{\rm B\kern-.05em{\sc i\kern-.025em b}\kern-.08em
    T\kern-.1667em\lower.7ex\hbox{E}\kern-.125emX}}

\usepackage[acronym]{glossaries}
\newacronym{sl}{SL}{STOP LEARNING}
\newacronym{inc}{INC}{INCREMENT}
\newacronym{dec}{DEC}{DECREMENT}
\newacronym{dac}{DAC}{Digital Analog Converter}
\newacronym{dpi}{DPI}{Differential Pair Integrator}    
\newacronym{ebn}{EBN}{Efficient Balanced Network}

\usepackage{hyperref}
\hypersetup{
    colorlinks=true,
    linkcolor=blue,
    filecolor=magenta,      
    urlcolor=blue,
}
    
\begin{document}

\bibliographystyle{unsrt} 

\title{Implementing efficient balanced networks with mixed-signal spike-based learning circuits}

\author{\IEEEauthorblockN{Julian B\"uchel, Jonathan Kakon, Michel Perez, Giacomo Indiveri}
\IEEEauthorblockA{\textit{Institute of Neuroinformatics} \\
\textit{University of Zurich and ETH Zurich}\\
Zurich, Switzerland \\
jubueche@ethz.ch}}

\maketitle

\begin{abstract}

  \glspl{ebn} are networks of spiking neurons in which excitatory and inhibitory synaptic currents are balanced on a short timescale, leading to desirable coding properties such as high encoding precision, low firing rates, and distributed information representation.
  It is for these benefits that it would be desirable to implement such networks in low-power neuromorphic processors.
  However, the degree of device mismatch in analog mixed-signal neuromorphic circuits renders the use of pre-trained \glspl{ebn} challenging, if not impossible.
  To overcome this issue, we developed a novel local learning rule suitable for on-chip implementation that drives a randomly connected network of spiking neurons into a tightly balanced regime.
  Here we present the integrated circuits that implement this rule and demonstrate their expected behaviour in low-level circuit simulations.
  Our proposed method paves the way towards a system-level implementation of tightly balanced networks on analog mixed-signal neuromorphic hardware.
  Thanks to their coding properties and sparse activity, neuromorphic electronic \glspl{ebn} will be ideally suited for extreme-edge computing applications that require low-latency, ultra-low power consumption and which cannot rely on cloud computing for data processing.
\end{abstract}


\glsresetall

\section{Introduction}
Many experiments have shown that cortical neural networks operate in a balanced regime, meaning that excitatory synaptic currents going into cortical neurons are balanced by their inhibitory counterpart on a fine time-scale~\cite{pmid8939866,PMID:15802011,pmid12748642}.
This enables such networks to have very low latency in response to sensory signals, to optimize their information coding properties, and to implement efficient distributed processing with sparse and low  spiking activity. 
However, the precise time-scales over which this balance is achieved is still highly debated~\cite{Ahmadian,hennequin,Deneve2016}. The \gls{ebn}, originally proposed in~\cite{Bourdoukan:2012:LOS:2999325.2999390}, is a network architecture derived under the condition that, in order to encode and represent a time-varying signal, no spike should be fired in vain, leading to a precisely balanced population code.
Although initially derived in an autoencoder framework~\cite{Bourdoukan:2012:LOS:2999325.2999390}, \glspl{ebn} can be augmented with a set of recurrent connections in order to carry out more complex operations, such as modeling non-linear dynamical systems~\cite{alemi2017learning}. Given their close link to cortical neural networks, and their favorable computational properties for processing sensory signals, \glspl{ebn} are an optimal computational primitive for neuromorphic electronic systems~\cite{Mead90,Chicca_etal14}.
However, as device mismatch can compromise performance, deploying pre-trained versions of these networks on mixed-signal neuromorphic hardware requires on-chip overhead or dedicated calibration procedures~\cite{Linares-Barranco_etal04,Neftci_Indiveri10}.

To address this issue we present a local discrete learning rule for on-chip training of recurrent spiking neural networks into \glspl{ebn}, that is optimally suited for mixed-signal neuromorphic circuits. We first validate the rule with high-level behavioural simulation, then we present the mixed-signal neuromorphic circuits that implement it and demonstrate their correct behaviour with low-level circuit simulations in two different experiments.

\section{Efficient Balanced Networks}
\subsection{Theoretical Background}
Let us assume that we have a population of $N$ spiking neurons into which $N_x$ continuous
signals are fed using a linear transformation $\mathbf{F}$. We furthermore assume that the population
of $N$ neurons is fully interconnected by a matrix $\mathbf{\Omega}$ and that we can reconstruct the continuous signal fed into the population using $\hat{x}_t = \mathbf{D}r_t$, where $r_t$ is the filtered spike train at time $t$ and $\mathbf{D}$ is a predefined decoding matrix so that $\mathbf{D} \propto \mathbf{F^T}$.
Following~\cite{Bourdoukan:2012:LOS:2999325.2999390}, we can derive the optimal network connectivity of the \gls{ebn} by assuming that neurons elicit spikes only when they contribute to reducing a loss function, which is mainly governed by the reconstruction error between the input $x$ and the reconstructed version, $\hat{x}$.   
Given a randomly chosen decoding matrix $\mathbf{D}$, the optimal feedforward ($\mathbf{F}$) and recurrent ($\mathbf{\Omega}$) weights are given by $\mathbf{D^T}$ and $\mathbf{-D^TD}$, respectively. For a full derivation please see \cite{Bourdoukan:2012:LOS:2999325.2999390} and the supplementary information of~\cite{wiel2017learning}. This leaves us with one equation describing the membrane potential  dynamics of the \gls{ebn}:
\begin{equation} \label{eq:dynamics}
    V(t) = \mathbf{F}x(t) + \mathbf{\Omega} r(t)
\end{equation}

Having introduced the optimal network connectivity of \glspl{ebn}, we can now derive the discrete learning rule to transform a randomly connected network into an \gls{ebn}. Instead of calculating the local gradient, as done by the learning rule
described in~\cite{wiel2017learning}, this rule checks whether a certain
condition is satisfied and increases or decreases the respective weights by a fixed discrete
step size.

Using the described optimal network connectivity and the assumption that the input can be reconstructed using $\hat{x}_t = \mathbf{D}r_t$, we can write the membrane potential of the $n$-th neuron to be
\begin{equation*}
  V_n(t) = D_n^Tx(t) - D_n^T\mathbf{D}r(t) = D_n^T(x(t)-\hat{x}(t))
\end{equation*}
Thus, minimising voltage fluctuations ensures that the projected reconstruction error is reduced.
This gives us the chance to minimise the simple loss function for all neurons $n$:

\begin{equation} \label{eq:loss}
  L = \frac{1}{2}(V_n^\textnormal{before}-V_n^\textnormal{rest} + V_n^\textnormal{after}-V_n^\textnormal{rest})^2
\end{equation}
where $V_n^\textnormal{before}$ is the $n$-th membrane potential \textit{before} the spike has propagated to the neuron and $V_n^\textnormal{after}$ is the potential after the spike has propagated. 
Similar to the way the optimal network connectivity was derived in~\cite{Bourdoukan:2012:LOS:2999325.2999390},
we impose the constraint that a recurrent connection should be changed by some minimal value $\omega$ if and only if it contributes to the reduction
of the loss. Therefore, on each spike that neuron $n$ receives, two conditions are checked:

\textbf{Condition I:}

\begin{equation*}
  L(\Omega_{n,k} + \omega) < L(\Omega_{n,k})
\end{equation*}

If we assume that the input signal does not change too rapidly, we can neglect the relatively small input term $\textnormal{dt}Fx(t)$ (see Eq. \ref{eq:dynamics}) and can write the voltage at neuron $n$ after a spike reached
this neuron as $V_n^\textnormal{after} = V_n^\textnormal{before} + \Omega_{n,k}$, if neuron $k$ emitted the spike.
Note that $\Omega_{n,k}$ is the recurrent connection from neuron $k$ to neuron $n$.
Using this relation, we plug in the definition of $V_n^\textnormal{after}$ into Eq. \ref{eq:loss} and reformulate to

\begin{equation*}
  L = \frac{1}{2}(2V_n^\textnormal{before} + \Omega_{n,k} - 2V_n^\textnormal{rest})^2
\end{equation*}

Plugging this into the condition with $\Omega_{n,k} + \omega$ on the LHS and $\Omega_{n,k}$ on the RHS yields

\begin{equation*}
  (2V_n^\textnormal{before} + \Omega_{n,k} + \omega - 2V_n^\textnormal{rest})^2 < (2V_n^\textnormal{before} + \Omega_{n,k} - 2V_n^\textnormal{rest})^2
\end{equation*}

\textbf{Condition II:} \\
The second condition is completely analogous to the first case with the only difference being that this time we check whether \textit{reducing} the weight by $\omega$ would reduce the loss: 

\begin{equation*}
  L(\Omega_{n,k} - \omega) < L(\Omega_{n,k})
\end{equation*}

After some reformulating (consult the repository for the full derivation), condition I and II can be summarised in the learning rule below:

\begin{equation} \label{eq:discrete_lr_hw}
  \Omega_{n,k} \leftarrow
  \begin{cases}
    \Omega_{n,k} + \omega \; \textnormal{if} \; 2V_n^\textnormal{before} + \Omega_{n,k} < -\frac{\omega}{2} + 2V_n^\textnormal{rest} \\
    \Omega_{n,k} - \omega \; \textnormal{if} \; 2V_n^\textnormal{before} + \Omega_{n,k} > \frac{\omega}{2} + 2V_n^\textnormal{rest} \\
    \Omega_{n,k} \; \textnormal{else}
  \end{cases}
\end{equation}

\subsection{Software simulations of the learning rule}

\begin{figure}
\centering
  \includegraphics[width = \columnwidth, height=6cm]{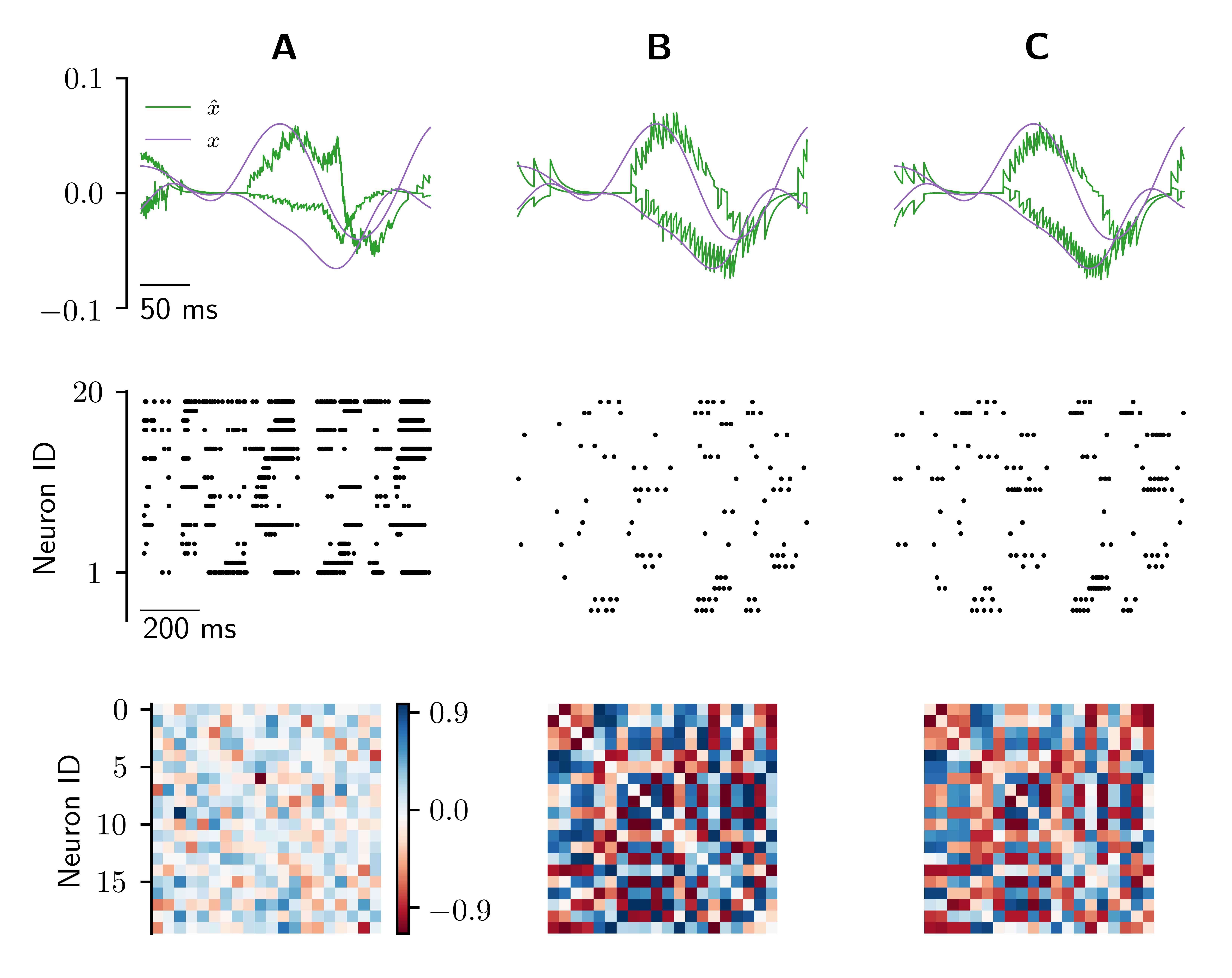}
  \caption{\textbf{A:} Network with random recurrent weights. After learning (\textbf{C}),
  the recurrent weights visibly match the optimal weights depicted in \textbf{B}. The code to run these simulations can be found \href{https://github.com/jubueche/On-Chip-EBN-Learning}{here}.}
  \label{fig:figure1}
\end{figure}

Given a random decoder $\mathbf{D}$, the recurrent connections of a balanced network
are pre-defined. In order to prove the correctness of our learning rule, it therefore suffices to check that, given a random $\mathbf{D}$, the initially random recurrent weights converge towards the optimal weights given by $\mathbf{\Omega} = \mathbf{-D^TD}$.
To validate this, we show an experiment where we simulate our learning rule in a population of 20 neurons for 50 iterations on a two-dimensional input that was generated by smoothing white noise with a Gaussian filter.
Figure \ref{fig:figure1} shows how the reconstruction accuracy improves dramatically, while at the same time reducing the average population activity and increasing sparsity (therefore reducing power consumption).
Note also how after only a few training iterations, the discrete recurrent matrix visibly approximates the real-valued optimal matrix, depicted in column B.
However, due to the discrete nature of the learning rule, with increasing levels of granularity the bounds imposed by the learning rule widen and fewer voltage deviations are penalised, leading to convergence towards non-optimal weight values.

\section{Hardware implementation}
\label{sec:hardw-impl}

\begin{figure}
\centering
  \includegraphics[width = \columnwidth]{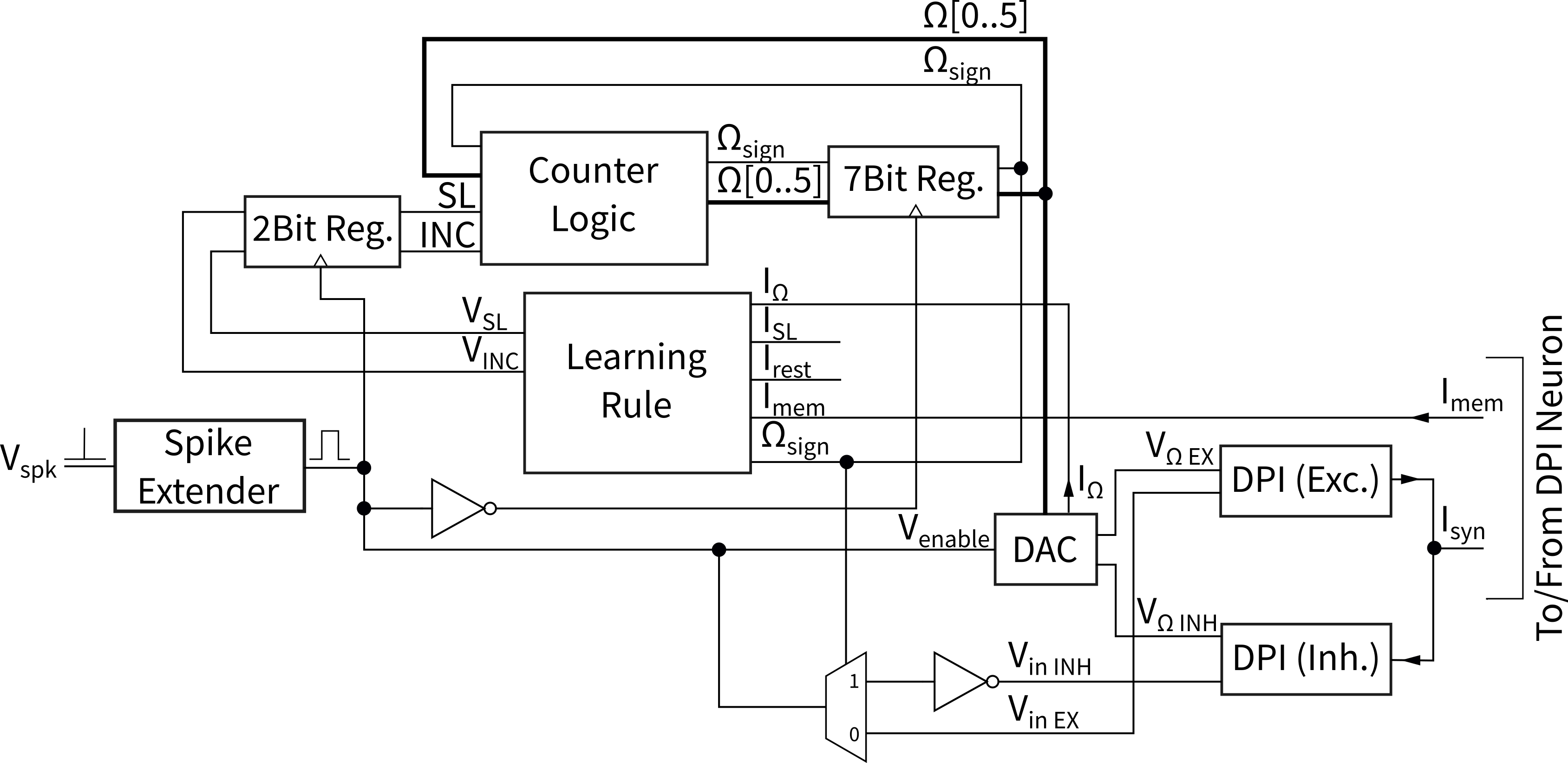}
  \caption{Overview of the hardware structure of a synapse. Biasing voltages and timing delays in the control signal are omitted. Variables used in the learning rule are displayed as currents. In the hardware implementations these are mirrored to the learning rule through diode connected MOSFETs.}
  \label{fig:overview}
\end{figure}

To implement the proposed learning rule in a neuromorphic processor, we need to be able to store the synaptic weight locally, compare it to the postsynaptic neuron membrane potential, update the synaptic weight, and source/sink the weighted current into/out of the neuron according to the saved state.
A general overview of the hardware structure is given in Fig.~\ref{fig:overview}. 
Input spikes, which typically last tens of nano-seconds, are processed by a spike extender circuit which creates pulses with variable pulse widths that can range from micro- to milli-seconds. This circuit is used to modulate the gain that the synapse has on the postsynaptic membrane potential. As the parameters of this circuit are shared across all synapses, this circuit can be used to effectively implement a synaptic scaling homeostatic mechanism to regulate the excitability of the neuron~\cite{Turrigiano99,Qiao_etal17}.
The extended pulse is then propagated to both the synapse integrator and the learning circuit.
Specifically, on the synapse integrator side it enables a current steering \gls{dac} to send the weighted post-synaptic current into a \gls{dpi} synapse circuit~\cite{Bartolozzi_Indiveri07a,Bartolozzi_etal06}, which then conveys the integrated synaptic current into the silicon neuron~\cite{Indiveri_etal11}. 
On the learning side, the rising edge of the input pulse triggers a two-bit register to store the digital state of the learning rule output set by the two control signals V$_\textnormal{INC}$ (for incrementing/decrementing the weight) and V$_\textnormal{SL}$ (for stopping the learning operation).
This is necessary since the learning rule reads the membrane potential of the neuron before the integrated synaptic current is injected into the silicon neuron.
The new weight is calculated in the ``Counter Logic'' block using the stored signals V$_\textnormal{INC}$ and V$_\textnormal{SL}$.
On the falling edge of the input pulse, the updated synaptic weight is finally stored into the seven-bit register, ready to be used when the next spike arrives.

\subsection{Current Steering Digital to Analog Converter}

\begin{figure}
\centering
  \includegraphics[width =\columnwidth]{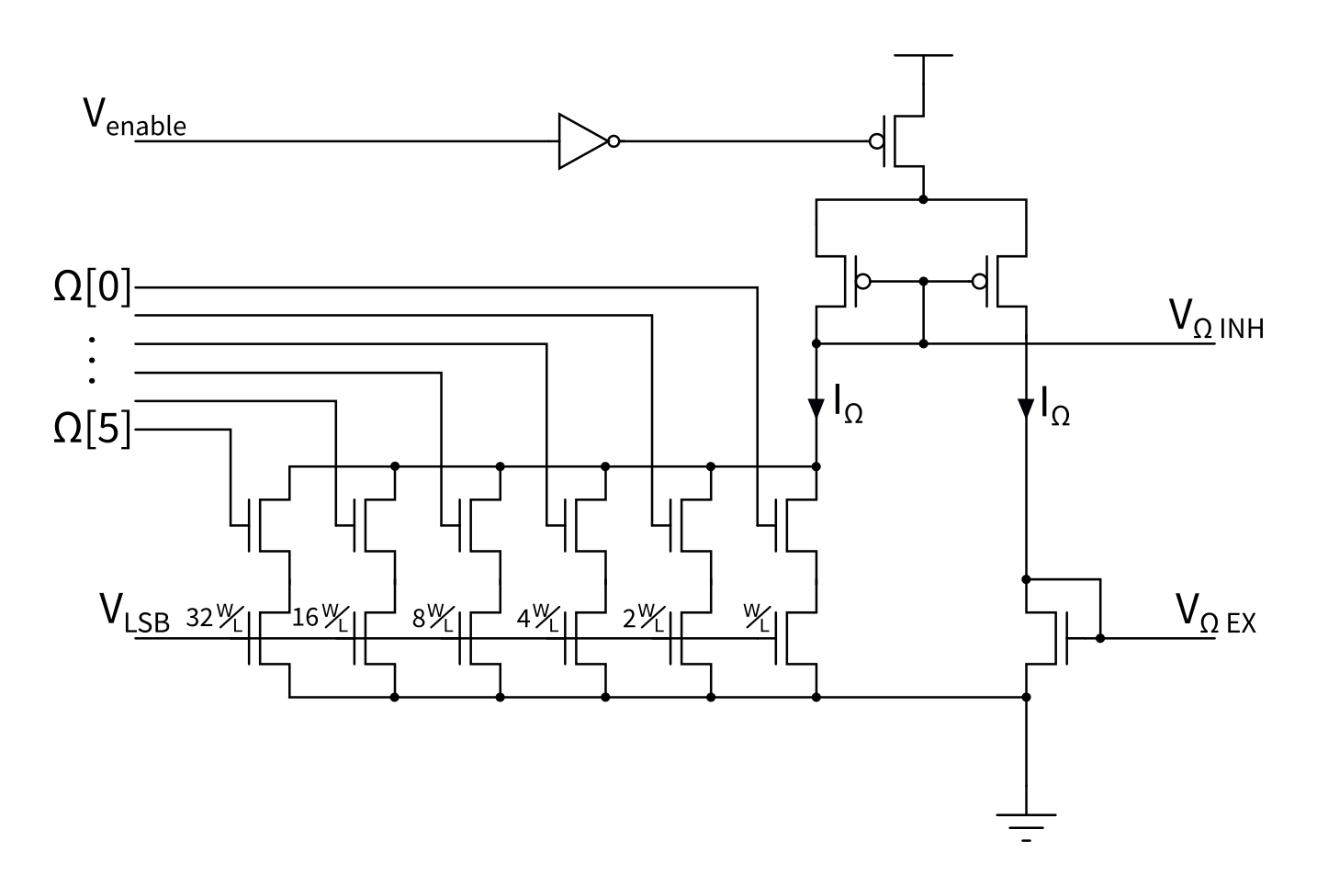}
  \caption{Binary weighted current steering \gls{dac}.}
  \label{fig:DAC}
\end{figure}

To convert the stored weight value into a current we use a binary weighted current steering \gls{dac} topology, where each bit is represented by two NFETs in series. The first acting as a switch, enabling the current path, and the second acting as a current sink scaling the current by the magnitude of the bit.
The summed currents are mirrored to a second complementary pair of transistors.
The two gate voltages of these transistors are used as the weight bias $V_{\Omega}$ of the excitatory and the inhibitory \gls{dpi}-synapse.
To save power, the \gls{dac} is enabled only for the duration of the extended input spike pulse.
Figure~\ref{fig:DAC} shows the schematic of the \gls{dac} used in our implementation.

\subsection{Learning Rule Module}

The learning rule module compares the current representing the post-synaptic neuron's membrane potential $I_{mem}$ with the presynaptic weight to evaluate the inequalities of \eqref{eq:discrete_lr_hw}.
If either one of the inequalities is satisfied, V$_\textnormal{SL}$ is set low and V$_\textnormal{INC}$ is set according to the decision whether to increment or decrement the weight. If neither of both inequalities is satisfied, V$_\textnormal{SL}$ is set high and no update is performed. By representing the weights and variables as currents we can rearrange the inequalities of eq.~\eqref{eq:discrete_lr_hw} to yield:

\begin{equation} \label{eq:lr_rearranged}
  \Omega_{n,k} \leftarrow
  \begin{cases}
    \Omega_{n,k} + \omega \; \textnormal{if} \; I_n^\textnormal{before} - I_n^\textnormal{rest} + \frac{I_{\Omega_{n,k}}}{2} + I_{SL} < 0 \\
    \Omega_{n,k} - \omega \; \textnormal{if} \; I_n^\textnormal{rest} - I_n^\textnormal{before} - \frac{I_{\Omega_{n,k}}}{2} + I_{SL} < 0 \\
    \Omega_{n,k} \; \textnormal{else}
  \end{cases}
\end{equation}

\begin{figure}
\centering
  \includegraphics[width = 1.0\columnwidth]{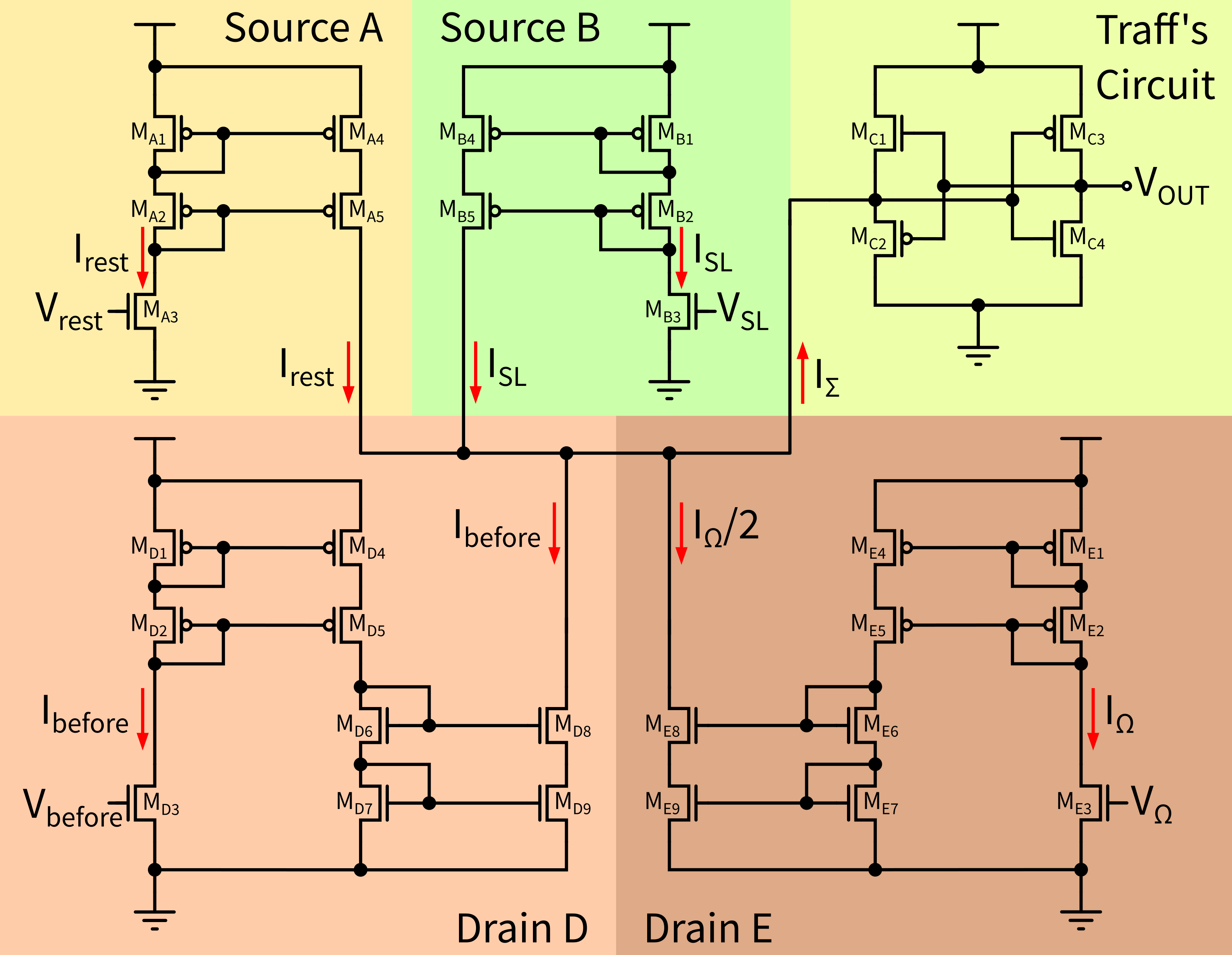}
  \caption{Circuit implementation for the evaluation of the second inequality of eq.\,\eqref{eq:lr_rearranged}. The bias voltages on $M_{A3}$ and $M_{B3}$ source the  $I_n^\textnormal{rest}$ and $I_\textnormal{SL}$ currents via the cascode mirrors \textit{Source A} and \textit{Source B} to Traff's circuit while \textit{Drain D} and \textit{Drain E} drain the currents  $I_n^\textnormal{before}$ and $I_{\Omega_{n,k}}/2$. When the inequality is satisfied $V_\textnormal{out}$ is pulled to $V_\textnormal{dd}$, and to \textit{GND} otherwise.}
  \label{fig:lr_schematic}
\end{figure}

Thanks to Kirchhoff's current law, the left side of the inequalities in eq.~\eqref{eq:lr_rearranged} can be easily realized by connecting the currents to a common node. Since the right sides of the inequalities are zero, we only have to check if the current sums are negative to determine if inequalities are satisfied or not. This is accomplished by the circuit illustrated in Fig.~\ref{fig:lr_schematic}.
The relevant currents are set by the bias voltages on $M_{A3}, M_{C3}$ and $M_{D3}$ which represent the input, while $M_{B3}$ is treated as a hyper-parameter for the learning circuit.
To reduce the Early effect and increase linearity, we make use of cascode current mirrors every time we need to mirror the currents.
Depending on the sign of the currents in the inequality, the currents are mirrored with either a sourcing (\textit{Source A, B}) or draining (\textit{Drain C, D}) configuration. All outputs of these current mirrors are connected to a common node, to calculate the value of $I_{\Sigma}$. We use a Traff current comparator ~\cite{Traff1992} to check the sign of $I_{\Sigma}$. This circuit is used once per inequality, therefore twice per synapse.

\section{Results}
As the circuits proposed operate in current-mode, all of the model variables are represented by currents. In the following, we indicate the membrane potential with $I_\textnormal{mem}$, and the neuron's reset and resting potential with $I_\textnormal{reset}$ and $I_\textnormal{rest}$ respectively.
To demonstrate the functionality of the learning rule we performed two basic experiments.

\begin{figure}
\centering
  \includegraphics[width = 1.0\columnwidth]{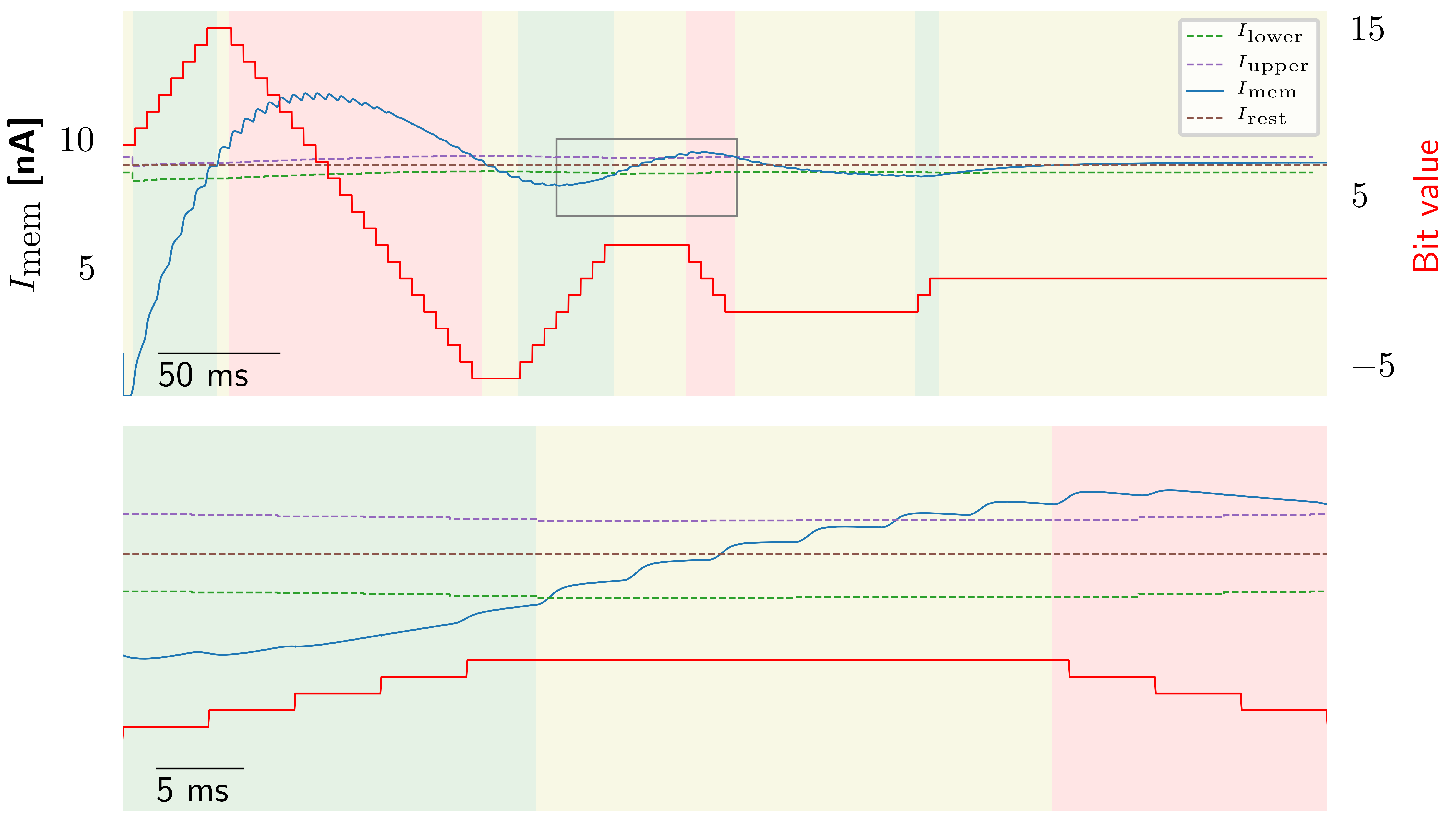}
  \caption{Changes in the weight value (red trace) and membrane potential state variable (blue trace) over time. The color of the shaded areas indicate either positive, negative or neutral correction of the synaptic weight. The upper and lower thresholds ($I_\textnormal{upper}$,$I_\textnormal{lower}$) around $I_\textnormal{mem}$ indicate the tolerance region within which the learning mechanism stops changing weights.}
  \label{fig:sim_1}
\end{figure}

In the first experiment, we train a single plastic synapse, stimulated with spikes at constant frequency, to balance the potential of a post-synaptic neuron around a nominal value $I_{\textnormal{rest}}$. %
In this experiment, we initially set the neuron's membrane potential to $I_{\textnormal{reset}}$ in order to simulate a state where the post-synaptic neuron was strongly inhibited. Since this is a large deviation from $I_{\textnormal{rest}}$ and the goal of the learning rule is to minimise fluctuations around $I_{\textnormal{rest}}$, the plastic synapse increases its weight having the effect of strongly and quickly increasing $I_\textnormal{mem}$ (see red and blue continuous traces in the first few ms of Fig.~\ref{fig:sim_1}). Due to the inertia of the learning rule, the neuron overshoots, bringing $I_\textnormal{mem}$ above $I_{\textnormal{rest}}$. At this point the learning circuit compensates for this positive deviation and starts to decrease the weight towards a negative value, making the neuron undershoot. Because of the negative feedback nature of the learning mechanism, this process converges to a steady state after a small number of oscillations around the target value where the weight stops changing.

\begin{figure}
\centering
  \includegraphics[width = 1.0\columnwidth]{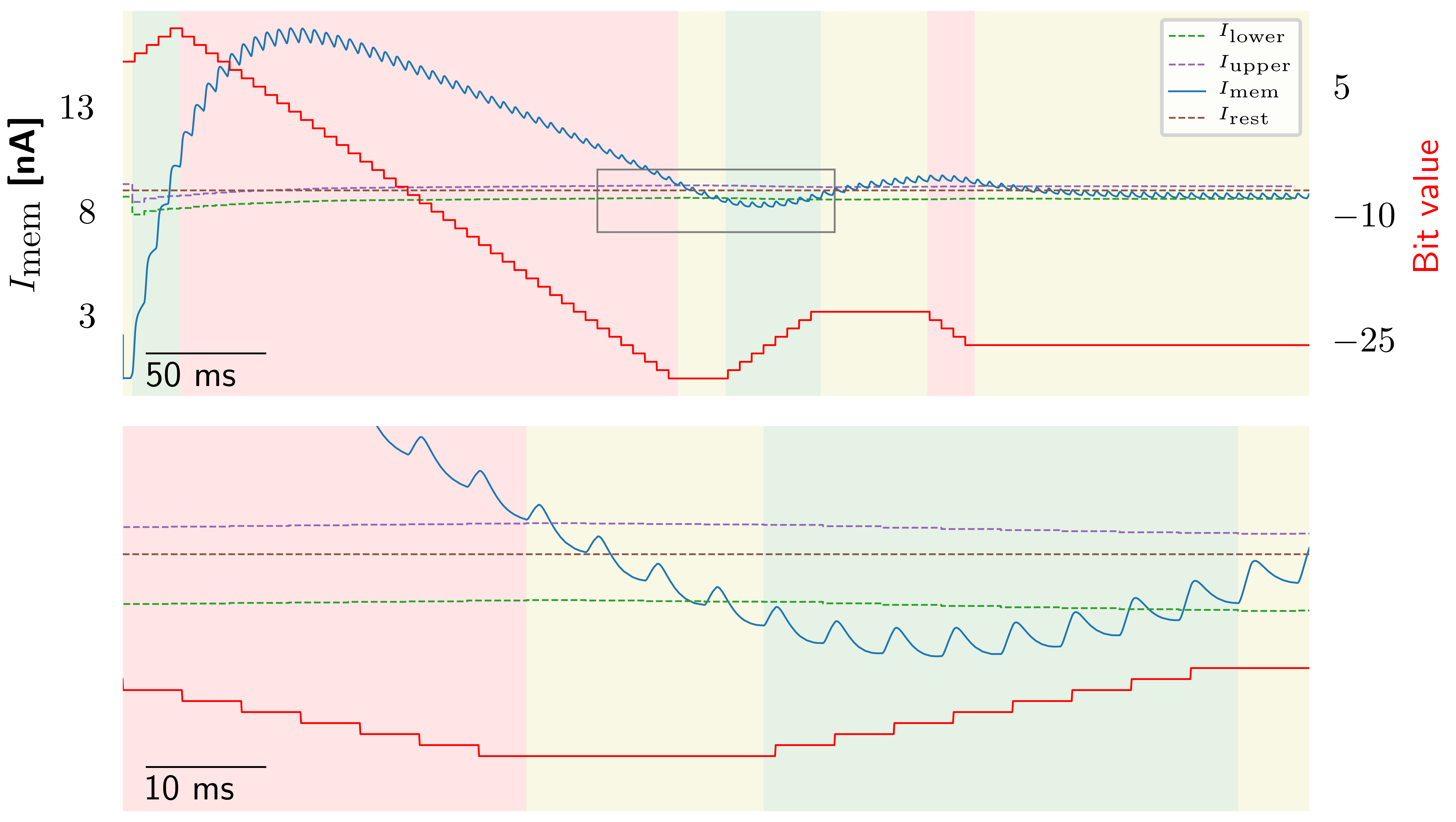}
  \caption{Changes of the membrane potential current and plastic weight value in the two-synapse stimulation experiment. The plastic weight increases or decreases following the rule of  eq.\,\eqref{eq:lr_rearranged} when $I_\textnormal{mem}$ does not fall within the circuit tolerance region.}
  \label{fig:sim_2}
\end{figure}

In the second experiment, the neuron receives input from two synapses: A plastic one and a fixed-weight excitatory one. Both synapses have the same time constant and are stimulated with the same spike train with a firing rate of 100\,Hz. As in the first experiment, we initialize the neuron to $I_{\textnormal{reset}}$, causing the learning rule to increase the synaptic weight. Figure~\ref{fig:sim_2} shows the response of the neuron to the two synaptic inputs, as the weight of the plastic synapse changes. Also, in this case the inertia of the learning rule causes the neuron to overshoot. This initial over-correction is then followed by a period of negative plastic weight updates which, compared to Fig.~\ref{fig:sim_1}, is much longer due to the effect of the non-plastic excitatory synapse. When the negative weight of the plastic synapse brings $I_\textnormal{mem}$ within tolerance bounds of $I_\textnormal{rest}$, the weight stops changing, and when $I_\textnormal{mem}$ falls outside of the tolerance region, the learning mechanism again adjusts the weights accordingly.

These experiments show that, in both cases, the synapse equipped with the learning rule learned to minimise the fluctuations around $I_\textnormal{rest}$, by either simply reducing its own weight or by balancing the injected current of another synapse in the system. These results demonstrate the correctness of our circuit implementation and, in combination with the network-level simulations validate the proper behaviour of the circuits for implementing neuromorphic electronic \glspl{ebn}.

\section{Conclusion}
We developed a novel learning rule that drives a randomly connected network of spiking neurons into a tightly balanced regime for producing \glspl{ebn}, and presented its hardware implementation using mixed-signal neuromorphic circuits.
We demonstrated the expected behaviour of the proposed learning rule with high-level behavioural simulations and the correct behaviour of its hardware implementation using low-level circuit simulations.
Future work will be devoted to the design of a prototype chip with the circuits presented and their integration into a large multi-core neuromorphic processor.
Further investigations will be carried out to assess the computations that can be done by such devices while harnessing the benefits of \gls{ebn}s (see for example~\cite{alemi2017learning,boerlin,eLife_opt_comp}).

\section*{Acknowledgment}
We are grateful to Christian Machens and members of the Machens lab for helpful discussions and their hospitality. We thank Arianna Rubino and Melika Payvand for providing guidance and support in the definition of the project and Mohammadali Sharifshazileh for helpful discussions and guidance with the circuit details. This paper is supported in part by the European Union's Horizon 2020 ERC project NeuroAgents (Grant No. 724295).

\bibliography{bibliography}

\end{document}